\documentclass[apjl]{emulateapj}
\usepackage{apjfonts}
\usepackage{color}

\newcommand{\hdust}[0]{{\sc hdust }}

\shorttitle{Modeling of IRAS\,00470+6429}
\shortauthors{A. C. Carciofi et al.}

\begin{document}

\title{Towards Understanding The B[e] Phenomenon: IV. Modeling of IRAS\,00470+6429.}

\author{A. C. Carciofi}
\affil{Instituto de Astronomia, Geof\'{i}sica e Ci\^encias Atmosf\'ericas, Universidade de S\~ao Paulo, 
Rua do Mat\~ao 1226, Cidade Universit\'aria, 05508-900, S\~ao Paulo, SP, BRAZIL}
\email{carciofi@usp.br} 

\author{A.~S. Miroshnichenko}
\affil{Department of Physics and Astronomy, University of North
Carolina at Greensboro, Greensboro, NC 27402, USA}
\and
\author{J. E. Bjorkman}
\affil{Ritter Observatory, M.S. 113, Dept. of Physics and Astronomy,
University of Toledo, Toledo, OH 43606-3390, USA} 

\begin{abstract}

FS\,CMa type stars are a recently described group of objects with the B[e] phenomenon that exhibit strong emission-line spectra and strong IR excesses.
In this paper we report the first attempt for a detailed modeling of IRAS\,00470+6429, for which we have the best set of observations.
Our modeling is based on two key assumptions: the star has a main-sequence luminosity for its spectral type (B2)  and the circumstellar envelope is bimodal, composed of a slowly outflowing disk-like wind and a fast polar wind. Both outflows are assumed to be purely radial.
We adopt a novel approach to describe the dust formation site in the wind that employs timescale arguments for grain condensation and a self-consistent solution for the dust destruction surface.
With the above assumptions we were able to reproduce satisfactorily many observational properties of IRAS\,00470+6429, including the \ion{H}{1} line profiles and the overall shape of the spectral energy distribution.
Our adopted recipe for dust formation proved successful in reproducing the correct amount of dust formed in the circumstellar envelope.
Possible shortcomings of our model, as well as suggestions for future improvements, are discussed.
\end{abstract}


\section{Introduction}\label{intro}

This series of papers is devoted to studying objects from the Galactic FS CMa type group. The group comprises about 40 objects with the B[e] phenomenon, which refers to the simultaneous presence of forbidden lines in the spectra and strong excesses of IR radiation \citep{m07,lam98}. Its members show properties of B-type stars with typical main-sequence luminosities [$\log(L/ L_{\sun}) \sim 2.5-4.5$]. On the other hand, they exhibit extremely strong emission-line spectra that are on average more than an order of magnitude stronger than those of Be stars of similar spectral types. The group includes objects that were discovered long ago, but their nature and evolutionary state was not satisfactorily determined, as well as newly found ones \citep{m07a}.

One of the main problems in understanding those objects has to do with the presence of significant amounts of hot dust in their circumstellar region that cannot be left from the time of the stars' formation. It is also hard to explain the dust formation using typical mass loss rates from single stars of an appropriate mass range ($\sim 5-20\;M_{\sun}$). Our observations show that over 30 per cent of the group objects show signs of secondary companions in the spectra (e.g., Li {\sc I} 6708 \AA\ absorption line). Therefore, one might assume that the large amounts of circumstellar matter near FS CMa type objects may be a consequence of the binary evolution, although direct evidence of the mass transfer between the stellar companions is not observed.

The main goals of this series of papers are to obtain high quality observational data for the group objects, to describe the objects' properties in detail, and to model the latter in order to put tight constraints on parameters of the stellar companions and those of the circumstellar material. The observed properties of the first object for which we collected a good wealth of data, IRAS\,00470+6429, were presented by \citet[][hereafter Paper III]{m09}. In the present paper we make use of the Monte Carlo code \hdust to determine the physical parameters of this intricate system.

\begin{table}
\begin{minipage}[t]{\columnwidth}
      \caption[]{Model Parameters for IRAS\,00470+6429}
	\label{t6}      
	\centering          
	\begin{tabular}{ l l l}  
	\hline\hline       
Parameter & Value & Type \\
\hline
\multicolumn{3}{c}{Stellar Parameters}\\ 
\hline
Spectral Type & B2V & fixed \\
$R$ & $6\;R_{\sun}$ & fixed \\
$T_\mathrm{eff}$ & $20\,000\;\rm K$ & fixed  \\
$L $ & $5\,100\;L_{\sun}$ & fixed \\
\hline
\multicolumn{3}{c}{Geometrical Parameters}\\ 
\hline
$i$ & $84\degr$ & free\\
$d$ & $1100 \pm 100 \rm\;pc$  & free\\
\hline
\multicolumn{3}{c}{Wind Parameters}\\ 
\hline
$d\dot{M}(90\degr)/d\Omega$ & $(1.4$ --- $1.6) \times 10^{-8}\;M_{\sun}\;\rm yr^{-1} \;sr^{-1}$ & calculated \\
$d\dot{M}(0)/d\Omega$ & $<3\times10^{-9}\;M_{\sun}\;\rm yr^{-1}\;sr^{-1}$ & free \\
$A_1$ & $>19$ & free \\
$v_\infty(90\degr)$ & $24\;\rm km\;s^{-1}$ & calculated \\
$v_\infty(0)$ & $1200\;\rm km\;s^{-1}$ & fixed \\
$A_2$ & -0.98& free \\
$\beta(0)$ & 4 & free \\
$A_3$ & 0 & free \\
$m$ & 92  & free \\
$R_\mathrm{w}$ &  $\gtrsim 10^5\;R_{\sun}$& fixed \\

\hline
\multicolumn{3}{c}{Common Dust Parameters}\\
\hline
Composition & Amorphous silicate \footnote{The optical constants of \citet{oss92} were used} & fixed \\
$n$ & -3.5 & fixed \\
Gas-to-dust ratio & 200 & fixed \\

\hline
\multicolumn{3}{c}{Dust Model 1}\\
\hline
$a_{\rm min}$\,---\,$a_{\rm max}$ & $0.05$\,---\,$0.25\;\rm \mu m$ & fixed \\
$\rho_{\rm dust}$ & $1.5\; \rm g\; cm^{-3}$ & free  \\
$T_{\rm destruction} $ & $1\,500\;\rm K$ & free  \\
$R_{\rm destruction} $ & $540\;R$ & calculated  \\
$\eta$ & 0.02 & free  \\
$E(B-V)_{\rm IS}$ & 1.1  & free\\
\hline
\multicolumn{3}{c}{Dust Model 2}\\
\hline
$a_{\rm min}$\,---\,$a_{\rm max}$ & $0.05$\,---\,$0.10\;\rm \mu m$ & fixed \\
$\rho_{\rm dust}$ & $0.15\rm\; g \;cm^{-3}$ & free  \\
$T_{\rm destruction} $ & $1\,500\;\rm K$ & free  \\
$R_{\rm destruction} $ & $540\;R$ & calculated  \\
$\eta$ & 0.02 & free  \\
$E(B-V)_{\rm IS}$ & 1.1  & free\\

\hline
\multicolumn{3}{c}{Dust Model 3}\\
\hline
$a_{\rm min}$\,---\,$a_{\rm max}$ & $1$\,---\,$50\;\rm \mu m$ & fixed \\
$\rho_{\rm dust}$ & $0.1\rm\; g \;cm^{-3}$ & free  \\
$T_{\rm destruction} $ & $1\,600\;\rm K$ & free  \\
$R_{\rm destruction} $ & $79\;R$ & calculated  \\
$\eta$ & 0.02 & free  \\
$E(B-V)_{\rm IS}$ & 1.2  & free\\
\hline
\end{tabular}
\end{minipage}
\end{table}

\section{Model Description}\label{modeling}

From the discussion of the available data in Paper III, it becomes clear that IRAS\,00470+6429 is a very complex object, with a large and dense CS envelope composed of both ionized gas and dust.
It is expected, therefore, that any model that successfully reproduces the observed properties must also be complex.

This first attempt of modeling of IRAS\,00470+6429 was conducted using the computer code \hdust \citep{car08b,car06a,car08a,car06b,car04}. \hdust is a three-dimensional Monte Carlo radiative transfer code that combines the full non-local thermodynamic equilibrium (NLTE) treatment of the radiative transfer in gaseous media with  a very general treatment for CS dust grains.
This code has been already successfully applied to study the CS disks of Be stars
\citep[see, for instance,][]{car07,car09}.

In {\sc hdust}, the CS environment can be represented by two components: a gaseous region, containing hydrogen and free electrons, and a dusty region. The solution to the problem is obtained in two steps. 
The first step consists of an iterative scheme to solve the coupled problems of radiative transfer, radiative equilibrium and statistical equilibrium.
For the gaseous shell, \hdust calculates the NLTE hydrogen level populations and the electron temperature as a function of position, taking into full account all the relevant physics (continuum and line transfer, collisional processes, etc.). 
For the dusty shells, \hdust solves the radiative equilibrium problem, thereby obtaining the equilibrium temperature of the dust grains as a function of position.
The interested reader is referred to  \citet{car06a} for details of the Monte Carlo NLTE solution and 
 \citet{car04} for information about the dust implementation.
Once the first step is completed and the state variables of the CS material are known, \hdust is run in a post-processing mode to calculate the emergent flux (spectral energy distribution, line profiles, synthetic images, etc.) for arbitrary lines of sight.

A few words about an important feature of the code  are warranted.
For a given gas plus dust distribution, as, for instance, the one adopted in Sect.~\ref{model}, \hdust solves for the gas and dust properties \emph{simultaneously and self-consistently}. Thus, important effects, such as the shielding of the dust grains from the stellar ionizing UV photons by the CS gas, are included in the calculations. 
This feature allowed us to determine for each model the \emph{dust destruction surface}, which is the latitude-dependent distance from star beyond which the dust equilibrium temperature is below an adopted destruction temperature.
This feature of the code proved to be a key factor for the 
model we describe below.


\subsection{A Parametric Model for the Circumstellar Envelope \label{model}}

For the problem at hand, \hdust must be provided with a description of the properties of the central star (effective temperature, $T_\mathrm{eff}$, and radius, $R$) and the physical properties of the CS envelope (density and velocity distribution of the gas and the characteristics of the CS dust, such as composition and grain size distribution). 
In addition, two geometrical parameters must be specified: the distance to the star, $d$, and the inclination angle, $i$, which is the angle between the observer's line of sight and the symmetry axis of the system. 
Finally, the interstellar reddening, $E(B-V)_{\rm IS}$,  must also be specified. Even though it is not an intrinsic parameter of the system, we include it as a model parameter because of the uncertainty in its determination, mainly owing to the difficulty to separate the interstellar from the circumstellar reddening.

A long standing issue regarding the stars with the B[e] phenomenon is that very little is known about the processes that lead to the formation of their CS environment \citep[see][for a review]{zic06}. Also, in spite of the effort of numerous observers and theoreticians, the most fundamental properties of the CS envelope remain quite uncertain.
In the case of supergiants B[e] (sgB[e]), a subclass of the stars with the B[e] phenomenon \citep{lam98}, it is generally believed that their CS envelope is bimodal, composed of a slowly outflowing, possibly rotating disk and a fast polar wind \citep{zic85}.
Two scenarios invoked to explain this bimodality are the rotationally induced bi-stability 
\citep{lam91} and rotationally supported Keplerian outflows \citep{por03}.


CS envelopes of FS CMa objects are even less understood, since very little theoretical work has been done with those objects so far. 
For this initial theoretical effort we adopt a model for IRAS\,00470+6429 that has two key assumptions. 
First, we assume that the star has a luminosity similar to that of a main-sequence star with the same spectral type, i.e., the stellar luminosity is in the range of a few thousand solar luminosities. We believe this assumption is well supported by the available data (Paper III).
Second, we assume that some sort of bimodal, two-component wind is present, similar in structure to the CS envelopes of sgB[e].
The results shown in Sect.~\ref{results} largely corroborate the above assumptions, but possible shortcomings are discussed in Sect.~\ref{discus}.

We adopt the following parametric form for the CS density and velocity structure.
We assume that the underlying physical reason for the bimodal envelope is that the mass loss is, somehow, enhanced around the equator. We express the stellar mass loss rate per unit solid angle as
\begin{equation}
\frac{d\dot{M}(\theta)}{d\Omega} = 
\frac{d\dot{M}(0)}{d\Omega}
\left[ 1 + A_1\sin^m(\theta) \right]\;,
\label{eq:mdot}
\end{equation}
where $\theta$ is the colatitude measured from the pole. $A_1$ is a parameter that controls the ratio between the equatorial
and polar mass loss rates
\begin{equation}
A_1 = \frac{d\dot{M}(90\degr)/d\Omega}{d\dot{M}(0)/d\Omega}-1\;.
\label{mdot}
\end{equation}
Since the mass loss rate is enhanced at the equator, $A_1 > 0$.
The parameter $m$ controls how fast the mass loss rate drops from the equator to the pole. Defining the {disk opening angle}, $\Delta\theta_{\rm op}$, as the latitude for which $d\dot{M}(\theta)/d\Omega$ has dropped to half of its equatorial value, we have
\begin{equation}
\Delta\theta_{\rm op} = \sin^{-1} 
\left[
\left(
\frac{A_1-1}{2A_1}
\right)^{1/m}
\right]\;.
\end{equation}

For the velocity structure, we assume that the gas motion is purely radial, with a latitude-dependent radial velocity given by
\begin{equation}
v_r(r,\theta) = v_0 + [v_\infty(\theta)-v_0](1-R/r)^{\beta(\theta)}\;,
\label{eq:velocity}
\end{equation}
where $r$ is the distance to the center of the star. 
For a given $\theta$, eq.~(\ref{eq:velocity})  is a standard $\beta$-law for radiatively driven winds 
\citep{lam99}.
We assume that both the wind terminal velocity, $v_{\infty}$, and the acceleration parameter, $\beta$, are functions of $\theta$, and have the same latitudinal dependence as the mass loss. Thus,
\begin{equation}
v_\infty(\theta) = 
v_\infty(0)
\left[
1+A_2\sin^m(\theta)
\right]\;,
\end{equation}
and
\begin{equation}
\beta(\theta) = 
\beta(0) 
\left[
1+ A_3 \sin^m(\theta)
\right]\;.
\end{equation}
As before, $A_2$ and $A_3$ control the ratio between the values of each quantity at the equator and at the pole,
\begin{equation}
A_2 = \frac{v_\infty(90\degr)}{v_\infty(0)}-1\;,
\end{equation}
and
\begin{equation}
A_3 = \frac{\beta(90\degr)}{\beta(0)}-1\;.
\end{equation}
Since  the terminal velocity at the equator is smaller than at the pole,
$A_2<0$.

Assuming radial outflow, the mass continuity equation is,
\begin{equation}
\frac{d\dot{M}(\theta)}{d\Omega} = r^2 \rho(r,\theta) v(r,\theta)\;,
\end{equation}
so we can write the CS gas mass density distribution as
\begin{equation}
\rho(r,\theta) = \frac{d\dot{M}(\theta)/d\Omega}{r^2 v(r,\theta) }\;.
\label{density}
\end{equation}

\subsection{Dust Condensation Site \label{dustrecipe}} 

Having chosen a prescription for the CS gas, we must now provide a site for dust formation and describe the properties of the dust grains that are formed.
In an initial study of the continuum emission of sgB[e], \citet{por03} used two criteria to determine the dust formation site. In his models, dust was formed where the gas temperature, which he assumed to be an ad hoc function of the distance from the star, was lower than a given value ($\approx1\,500\;\rm K$) and the gas density was larger than a critical value. 

In our models, the coolest part of the envelope is in the denser equatorial regions, an expected result based on previous works \citep[e.g.][see Fig.~\ref{temperature}, below]{zsa08,car06a}.
However, even in the midplane the gas temperatures never fall below about $4\,000\;\rm K$. This temperature is much larger than the temperature needed for dust grain condensation to occur, which is thought to be below about  $1\,500\;\rm K$ \citep{por03}. 
It is likely that the gas temperatures predicted by \hdust are overestimated, because no cooling 
by helium or other metals is included. 
However, \citet{zsa08}, in a study of the properties of the CS envelope of sgB[e], found that the temperature never falls below $2\,000$ --- $3\,000\; \rm K$, even though they used more realistic cooling terms. These authors argue that a possible mechanism for further cooling the gas is adiabatic cooling, or an even more complex chemical composition.

Because of the above, it is clear that the gas temperature cannot be used in our models to trace
 the dust formation site and we use, for this purpose, a different approach than the one used by \citet{por03}.
 We define the dust formation site using two complementary criteria. 
If dust is to be present at a given location of the CS envelope,
 
i) The equilibrium temperature of the dust grains (not the gas temperature) must be smaller than a given grain destruction temperature, $T_{\rm destruction}$. 

ii) Furthermore, the gas density must be larger than a critical value, such that grain growth can occur.

The first criterion was implemented in \hdust using a very simple procedure.
For a given model, we start from a guess of the distance from the star beyond which the dust grains begin to condensate. 
During the iterative process used for determining the temperatures, whenever the dust temperature in a given cell becomes larger than $T_{\rm destruction}$, the code removes the dust from that particular cell. 
Conversely, if in a given dust-free cell the temperature of the dust, if present, would have been smaller than the destruction temperature, then that cell becomes a dusty cell.
This simple procedure ensures a self-consistent determination of the location of the dust destruction radius as a function of latitude.
Note that the dust destruction radius is highly model dependent: it is mainly controlled by the gas opacity in the inner part of the envelope and the dust model used.

The second criterion for defining the dust formation site follows a simple argument outlined by \citet{gai88} and adopted by \citet{por03}. 
The idea is that a critical gas number density for the dust formation can be derived by comparing two timescales, the timescale for chemical reactions, $\tau_{\rm ch}$,  and the timescale for the gas expansion, $\tau_{\exp}$. Grain formation and growth occurs if $\tau_{\rm ch} < \tau_{\exp}$. \citet{gai88} found that this occurs if the number density of the growth species (silicon, in this case) is larger than the following critical value
\begin{equation}
n_{\rm Si}(r,\theta) = \frac{v_r(r,\theta)}{10^{-16} r v_{\rm rel}(r,\theta) } \;,
\label{rho_crit}
\end{equation}
where $v_{\rm rel}$ is the relative velocity between the dust grain and an atom or molecule, which is of the order of the sound speed and, thus, position-dependent.
Since $n_{\rm Si}$ of eq.~(\ref{rho_crit}) was defined from simple physical arguments, and, therefore, is probably just a rough estimate of the critical density, we introduce a scaling factor, $\eta$, 
so that the condition for dust formation becomes
\begin{equation}
\frac{\epsilon \rho(r,\theta)}{\mu m_{\rm H}} \le \eta n_{\rm Si}(r,\theta)\;,
\label{rho_crit2}
\end{equation}
where $m_{\rm H}$ is the mass of the H atom and $\mu$ is the gas molecular weight, which is typically about 0.6 for an ionized gas with solar abundance.
$\epsilon$ is the relative abundance in number of Si to H, for which we adopt the solar value $3.6\times10^{-5}$.
The IR excess is a measure of the amount of dust that is formed in the CS envelope. Hence, the value of $\eta$ can be found by matching the IR excess.
The second criterion for the spatial distribution of the CS dust effectively confines the dust grains to the dense and slow equatorial regions.

Finally, given a site for the dust in the CS envelope, we need to describe the properties of the dust grains, which requires several additional parameters.
We assume that at each location of the envelope where dust can form, a certain fraction of the gas mass is converted into dust. This fraction, known as the gas-to-dust ratio, was assumed to be 200.
We further assume that grains are spherical and their composition is oxygen-based, i.e., the dust is comprised mainly of silicates. This assumption is supported by evolutionary arguments, since carbon is depleted due to the CNO processing cycle in massive stars, therefore ruling out carbon-based grains, such as graphite, for the object in question.
Furthermore, IR spectra of nearly 20 FS CMa objects obtained with the Spitzer Space Observatory (IRAS00470+6429 was not observed due to the presence of a bright nearby IR source) show silicate emission features \citep{m08}.

For this initial analysis we adopt a MRN law for the distribution of grain sizes \citep{mat77}, according to which
\begin{equation}
f(a) = C a^n\;, \label{MRN}
\end{equation}
where $f$ is the fractional number of dust grains with radius $a$ and $C$ is a normalization constant.
In the standard MRN law, $n=-3.5$ and the distribution is confined
between $a_{\rm min} = 0.05\; \mu\rm m$ and $a_{\rm max} = 0.25\; \mu\rm m$. In our models, we let 
both $a_{\rm min}$ and
$a_{\rm max}$ to be free parameters in order to investigate the possible existence of large dust grains (Paper III).
Given a dust mass density and $f(a)$, we further need the average density of the grain material, $\rho_{\rm dust}$, in order to calculate the number density of dust grains. Since we do not know how fluffy the grains are, we let $\rho_{\rm dust}$ be a free parameter.
We note, however, that $\rho_{\rm dust}$ and the gas-to-dust ratio are \emph{not independent parameters}, since both are contained in the dust optical depth, the one quantity that effectively controls the radiative transfer in the CS dust shell. 

\subsection{Adopted Grid Structure}

A few words are needed about the grid structure we adopted for the current problem.
In {\sc hdust}, the CS envelope is divided into $N_r$ radial cells, $N_\mu$ colatitude cells and $N_\phi$ azimuth cells, and it is assumed that the state variables of the CS material (temperature and level populations) are constant in each cell. 

The density law of eq.~(\ref{density}) can present difficulties for a radiative transfer solver, because it has a large radial gradient close to the star and it may have a sharp enhancement at the equator, depending on the adopted value of $m$.
For the radial spacing we adopt cells sizes such that the radial electron scattering optical depth is the same for all cells, which results in very small cells close to the star and large cells in the outer envelope. This ensures that the inner regions of the envelope are sampled adequately.
For the colatitude bins, we define the cell sizes is such a way that from the equator to the pole, along a ray with constant radius, the density at the center of consecutive cells falls in steps of $2/N_\mu$.
Therefore, when $m$ is large, which corresponds to a disk with small opening angle, the cells are small at the equator and large at the pole.

Two competing issues must be taken into account when choosing the values of $N_r$ and $N_\mu$.
Since \hdust employs a statistical method for sampling cell-dependent quantities such as radiative heating \citep[see][]{car06a}, the larger the number of cells the larger the computational effort necessary to reach a given sampling error in the rates.
On the other hand, the grid structure must be fine enough to describe correctly the adopted density and to resolve possible regions of ionization changes, where a transition from ionized to neutral hydrogen occurs.
The minimum number of cells necessary for a given problem can be estimated by running models with an increasing number of cells. If the cell spacing is properly defined, there will be a certain number of cells beyond which no modification of the solution is found by further refining the grid.
We found that $N_r=100$ (60 for the gas shell and 40 for the dusty shell) and $N_\mu=40$ are a good compromise between speed and accuracy.
For the current problem the number of azimuth cells was set to 1 because axisymmetry was assumed.

\subsection{Modeling Procedure \label{modeling procedure}}

The adopted model described above has many parameters but for this initial study we chose to let just part of them be free parameters, while keeping the rest fixed.
We chose as fixed parameters the ones which can either be sufficiently well-constrained by a model independent analysis, or are associated with the basic model assumptions defined in Sect.~\ref{model}. Examples of fixed parameters are the stellar characteristics and the dust composition.

Free parameters, on the other hand, are the ones about which no information can be readily obtained from the observations and, therefore, require a detailed modeling. The main free parameters in our study describe the density and velocity of the CS gas and the latitude-dependent stellar mass loss rate.
We believe that this approach is appropriate for this initial study, since it allows us to focus on the least known properties of the CS envelope.
A list of model parameters for IRAS\,00470+6429 is given in Table~\ref{t6}.

\subsection{Stellar Parameters \label{stel_par}}

The stellar parameters are all fixed in this study.
As mentioned above, one of our key assumptions is that  IRAS\,00470+6429 has a typical luminosity of a normal main-sequence B star. Following the analysis in Paper III we chose a spectral type of B2V for the central star. This sets the radius to $\approx6\;R_\sun$, the effective temperature to $\approx20\,000\rm\;K$ and the luminosity to  $\approx5\,100\;L_\sun$ \citep{har88}.
We investigated the effects of changing $T_{\rm eff}$ and $L$, and we found that varying $T_{\rm eff}$ by $\approx10$\% (and thus $L$ by $\approx40$\%) has only small effects on the results. For this initial study we considered a static, non-rotating central star. This is not a limitation of \hdust \citep[see, e.g.][]{car08a} but, instead, reflects the fact that the observed line profile is controlled mostly by the CS gas (Sect.~\ref{CSgas}).

For the stellar spectrum we used the model atmosphere of \citet{kur94} for $T_{\rm eff} = 20\,000\rm\;K$ and $\log g = 4$ \footnote{The use of a LTE model atmosphere is not a limitation of the present work. More recent NLTE results could have been used, but the object is so deeply embedded in the CS outflow that differences in the emerging flux would be negligible (see Sect.~\ref{CSgas}). }. To describe the photospheric specific intensity as a function of direction we adopted the limb darkening models of \citet{cla00}.

\subsection{Geometrical Parameters}

The geometrical parameters ($i$, $d$ and interstellar reddening) 
are all free parameters.  Nothing is known about $i$, but our previous analysis allows us to impose some limits on $d$ and $E(B-V)_{\rm IS}$. Following Paper III, we adopt $d$ in the range of 1000 --- 2000 pc and  $E(B-V)_{\rm IS}$ in the range 0.8 --- 1.2 mag.
Given a model for the CS gas and dust, $d$ and $E(B-V)_{\rm IS}$ are obtained by fitting the observed flux level and shape of the visible SED, respectively.

\section{Results \label{results}}

\begin{figure}[!t]
\begin{center}
\includegraphics[width=8cm]{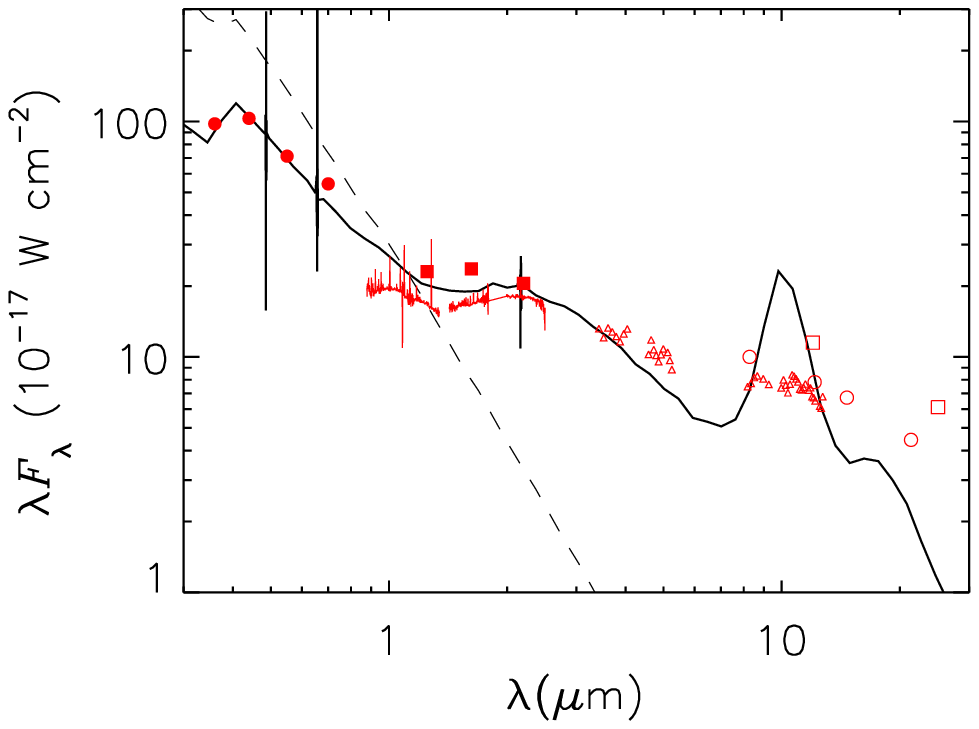} \\
\includegraphics[width=8cm]{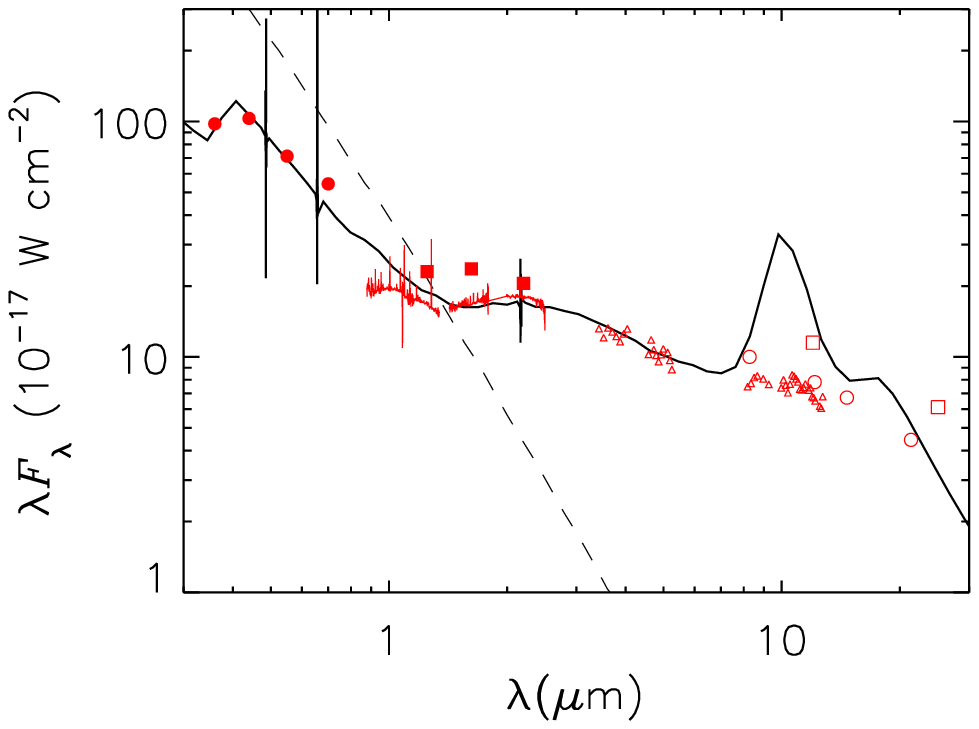} \\
\includegraphics[width=8cm]{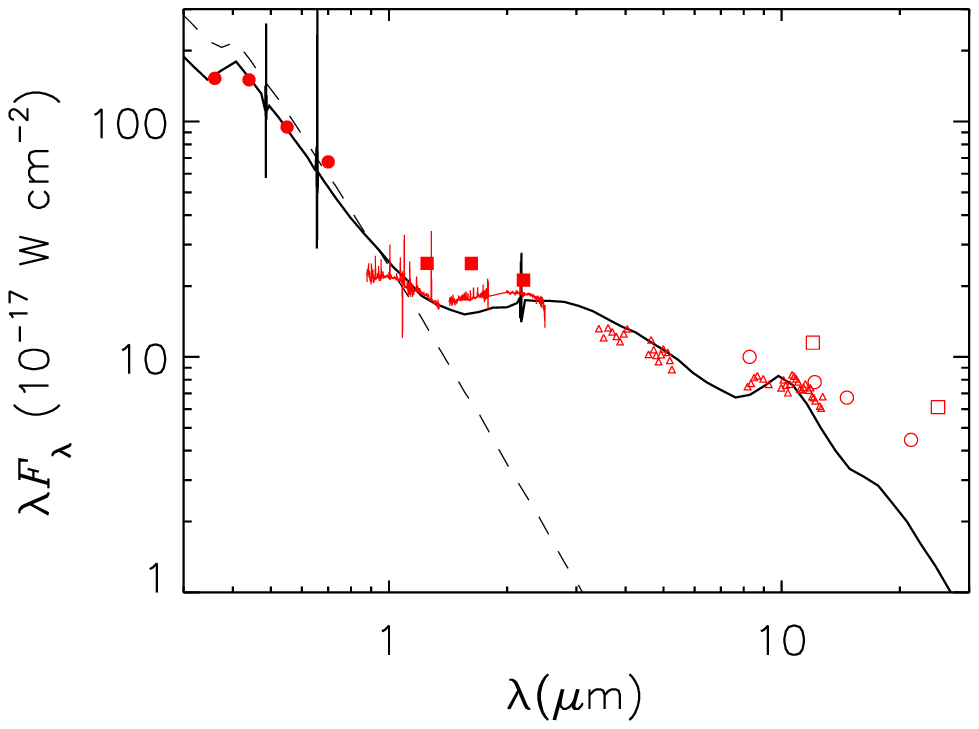}
\end{center}
\caption{
Comparison between observations and the model SED for the three dust models used.
The observations (red lines and symbols) are described in Paper III. 
{\it Symbols}: circles -- $UBVR$ photometry, squares --
near-IR photometry from 2MASS, triangles -- BASS data, large open circles -- MSX data,
large open squares -- IRAS data, and solid lines -- 2003 NIRIS
data. 
{\it Top:} Dust model 1. 
{\it Middle:} Dust model 2. 
{\it Bottom:} Dust model 3. 
In all panels the dashed line represents the unattenuated stellar spectrum.
Vertical lines on the best-fit model SED are hydrogen emission lines (from left to right -- H$\beta$, H$\alpha$, and Br$\gamma$).
\label{sed1}}
\end{figure}

\begin{figure*}[!t]
\centerline{\plottwo{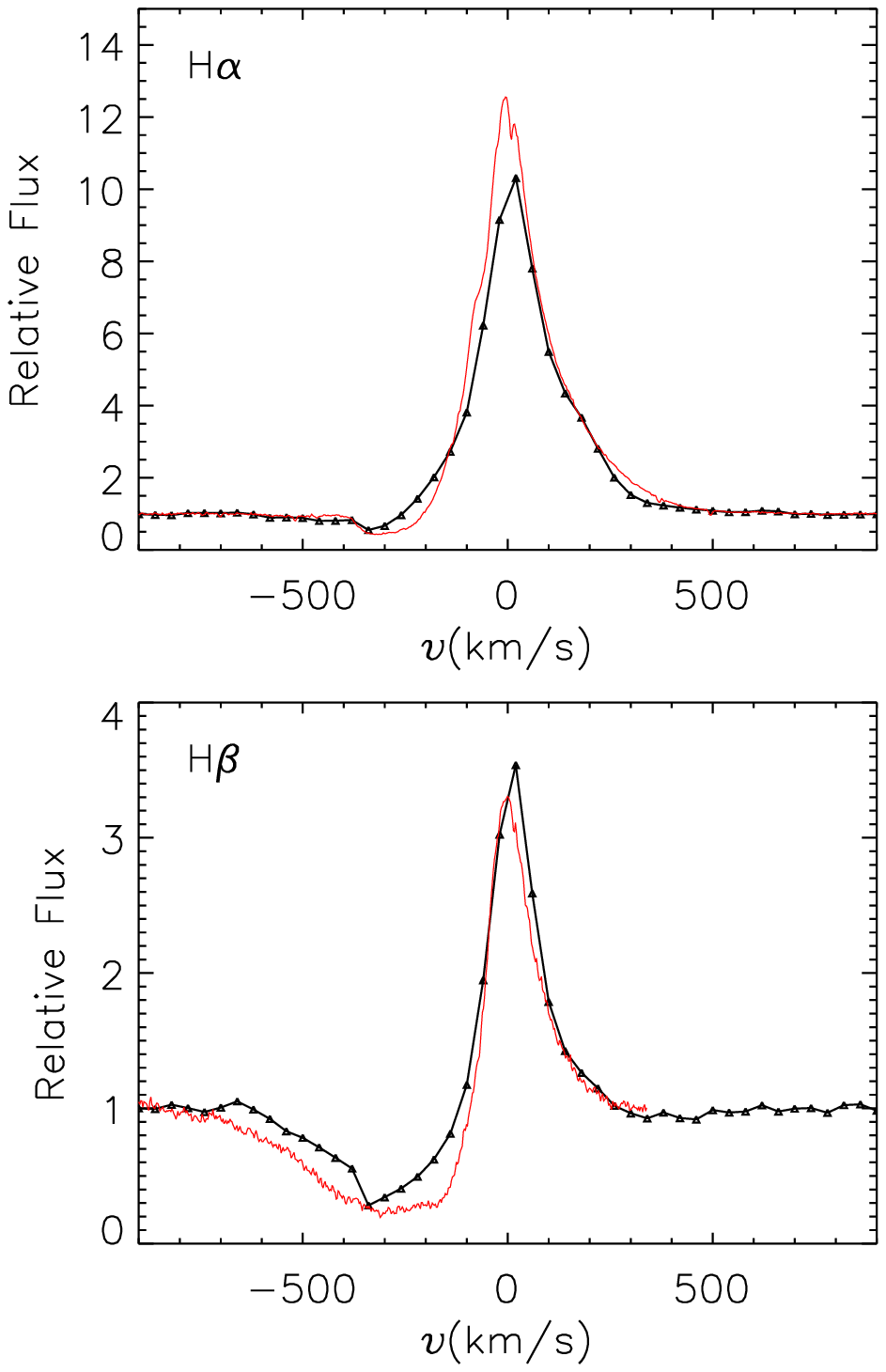}{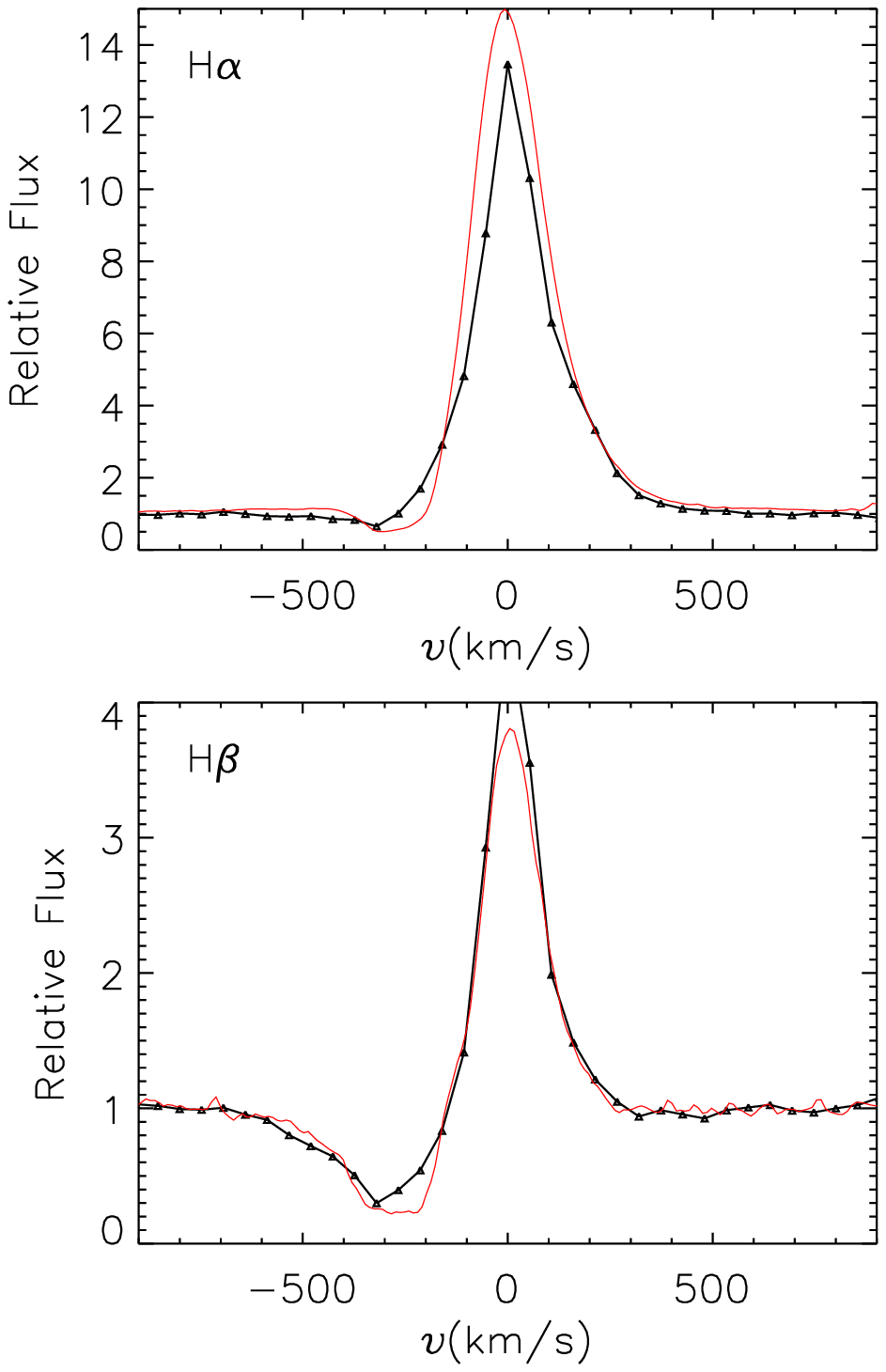}}
\caption{
Comparison between the model (black) and observed (red) line profiles. The observations and instruments used are described in Paper III.
{\it Left:} Best fit for the 2006, December 13 epoch, using $d\dot{M}(90\degr)/d\Omega=6 \times10^{-9}\;M_\sun \rm\;yr^{-1}\;sr^{-1}$.
{\it Right:} Best fit for the 2006, December 27 epoch, using $d\dot{M}(90\degr)/d\Omega=1\times10^{-8}\;M_\sun \rm\;yr^{-1}\;sr^{-1}$.
\label{hahb}}
\end{figure*}

\begin{figure}
\centerline{\includegraphics[width=8cm]{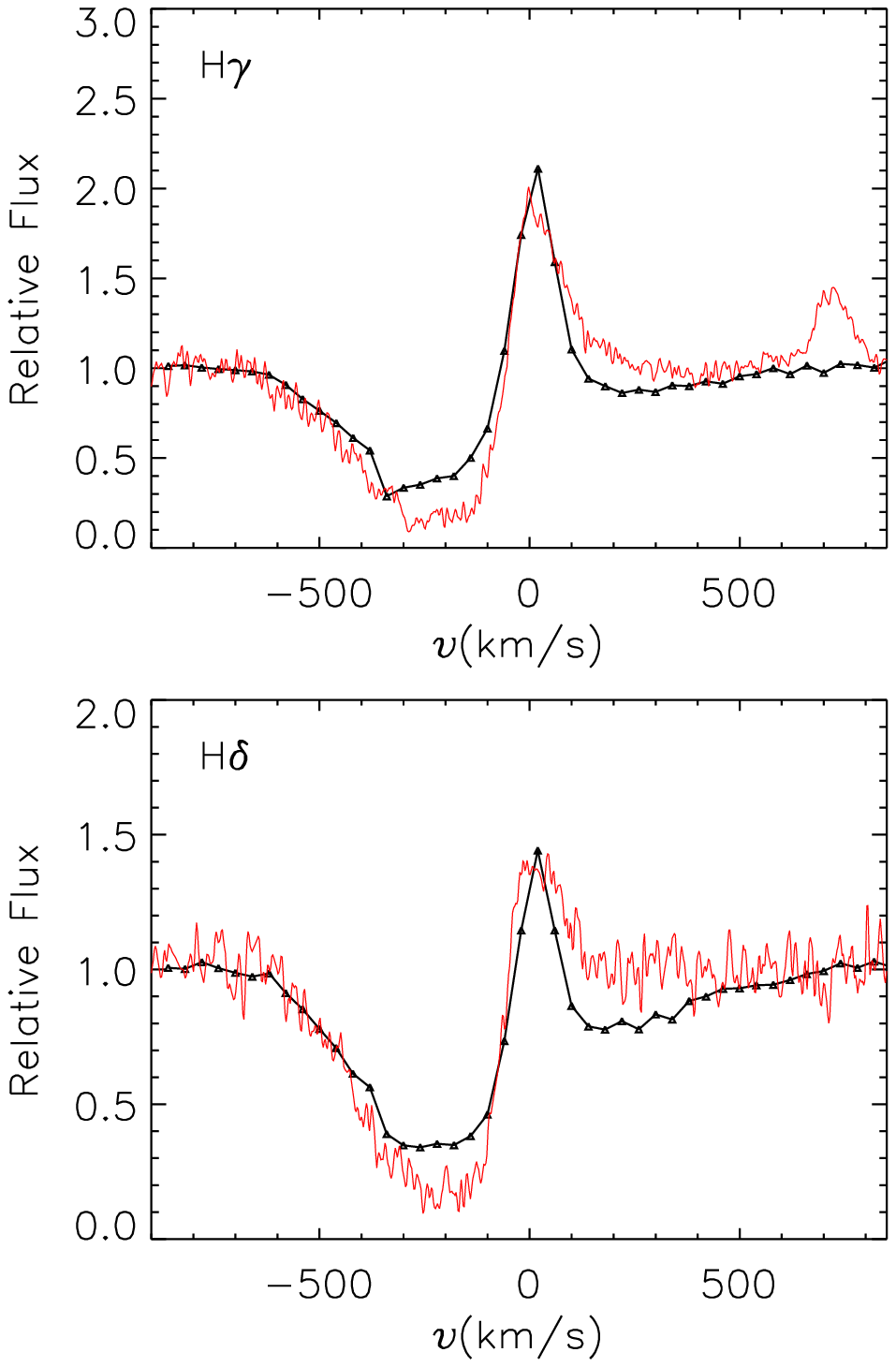}}
\caption{Comparison between the observed H$\gamma$ and H$\delta$ line profiles (2006, December 13th epoch, red) and models (black).
\label{hghd}}
\end{figure}

\subsection{Circumstellar Gas \label{CSgas}}

Some information about the gas dynamics can be readily obtained by analyzing the emission line profiles.
For instance, the broad blue absorption profile of the \ion{H}{1} lines suggests that the absorbing material has large velocities in the line of sight and the blue edge of the absorption profile indicates the terminal velocity of the fast wind. 
We adopted for $v_\infty(0)$ a value of {$1200\;\rm km\;s^{-1}$} and set this as a fixed parameter in our modeling. 
We note that this value of the terminal velocity is consistent with what is expected for main-sequence B stars \citep{gra87}.

In addition, the relatively narrow peaks of the Balmer lines suggest that the bulk of the emitting material has small velocity along the line of sight. Supposing that most of the \ion{H}{1} line emission comes from the equatorial region (an assumption corroborated by our detailed calculations) this suggests that the equatorial outflow velocities are very small compared to the velocities in the polar region. 

For the present work, hundreds of models were run covering a wide range of model parameters.
Before presenting our solution, it is useful to discuss some trends that became clear once a large range in the parameter space was covered.

From our results we can safely rule out both large 
($\Delta\theta_{\rm op} \gtrsim 15 \degr$ or $m \lesssim 15$) and small ($\Delta\theta_{\rm op} \lesssim 5 \degr$ or $m \gtrsim 180$) opening angles for the outflowing disk around IRAS\,00470+6429.
The opening angle, together with the equatorial mass loss rate, controls the strength of the  \ion{H}{1} line emission. 
For large opening angles, relatively low mass loss rates are necessary to reproduce the observed emission strength, i.e.,  the larger the opening angle, the lower the equatorial densities for a given emission strength.
Also, the lower the density, the farther away from the star is the dust destruction surface and, therefore, the lower the density in the dust condensation zone.
The two effects combined have the consequence that  models with $\Delta\theta_{\rm op} \gtrsim 15 \degr$ have densities below the critical density necessary to form dust  (eq.~[\ref{rho_crit2}]).

Models with small opening angles require high equatorial densities to reproduce the line emission and, therefore, can form dust.
 However, as we shall see in Sect.~\ref{CSdust}, in our models dust is typically confined within a wedge about one opening angle away from the equator.
This is because the density drops quickly towards the pole and soon falls below the critical density for grain formation.
If the opening angle is too small ($\Delta\theta_{\rm op} \lesssim 5 \degr$), the dust will encompass too small a solid angle for the radiation that comes from the inner envelope (both stellar radiation and reprocessed radiation by the CS gas) and, as a consequence, will reprocess little radiation. This is a geometrical effect and is true even for very large dust opacities.
Therefore, small opening angles can be ruled out on the grounds that they produce too small IR excess.

Another trend that became evident is that $d\dot{M}(0)/d\Omega$ cannot be larger than about $3\times10^{-9}\;M_\sun\;\rm yr^{-1}\;sr^{-1}$.
A large density in the poles implies that the polar regions would contribute significantly to the overall line emission. Since the polar material has a large terminal speed, the polar emission would appear as a broad red emission wing, which is not observed.
The polar emission begins to alter the emission profile
significantly if mass loss rate exceeds the quoted value.

This upper limit for $d\dot{M}(0)/d\Omega$ has an important implication on the possible range of inclination angles. Since the polar regions must be relatively optically thin, they cannot account for the broad and deep absorption component observed in the Balmer lines (see, for instance, Fig. 3 of Paper III).
Models with $0 < i \lesssim 75\degr$ do not produce the necessary absorption and, therefore, this range of inclination angles can be excluded. Besides this interval, our models can also exclude inclination angles too close to edge-on ($i \approx 90\degr$).
In all our models the radial optical depth along the equator is very large ($\tau \approx 10^6$ in the Lyman continuum and $\approx 10^2$ in the Balmer continuum). As discussed in Paper III, some features of the stellar photosphere are observed, which, therefore, excludes edge-on viewing.


Our best-fit SED, H$\alpha$, H$\beta$, H$\gamma$ and H$\delta$ line profiles for the parameters listed in Table~\ref{t6} are compared to the observations in Figs.~\ref{sed1} to~\ref{hghd}.
The CS gas density and velocity profiles along several latitudes are shown in Fig.~\ref{density}.

As said above, the opening angle and the equatorial mass loss rate are the primary parameters that control the strength of the line emission for a given $i$.
Our best-fit value for the opening angle is  $7\degr$, which corresponds to $m=92$.
For the equatorial mass loss rate our best-fit value is in the range $d\dot{M}(90\degr)/d\Omega=(1.4$ --- $1.6)\times10^{-8}\;M_\sun \rm\;yr^{-1}\;sr^{-1}$.
The lower limit corresponds to the best-fit value for the line emission observed in 2006 December 13, and the upper limit reproduces better the 2006 December 27 observations, which has substantially stronger emission (see Fig.~\ref{hahb}).
This result indicates that IRAS\,00470+6429 presents significant changes in the mass loss rate in very short timescales (two weeks).

Both $m$ and  $d\dot{M}(90\degr)/d\Omega$ are well constrained parameters in our models, but we note that reasonable fits of the emission strength could also be obtained with slightly larger opening angles (say, by one or two degrees) and slightly smaller (a few tens of per cent) $d\dot{M}(90\degr)/d\Omega$. The opposite is also true, i.e., similar fits can be obtained by slightly smaller opening angles and slightly larger mass loss rates.

The opening angle and the mass loss rate also control the SED shape. Their effects on the SED are complex, and depends significantly on the adopted $i$. Dust also plays a major role in defining the shape of the SED, not only in the IR but also on the visible spectral region provided that the line of sight of the observer crosses dusty material. For this reason, we postpone the discussion of the model SED and its comparison to observations to Sect.~\ref{CSdust}.

Contrary to the equatorial mass loss rate, the polar mass loss rate is rather poorly constrained. As explained above, an upper limit of about $d\dot{M}(0)/d\Omega=3\times10^{-9}\;M_\sun \rm\;yr^{-1}\;sr^{-1}$  could be set on the basis of the absence of a broad emission component. 
Given the best-fit value of $d\dot{M}(90\degr)/d\Omega$, above, this corresponds to a lower limit of $A_1=19$, which means that the ratio between the equatorial and polar mass loss rates is at least as high as 20, but our models do not exclude much larger values of $A_1$. 
The reason why our models do not constrain well the polar densities is that the \ion{H}{1} recombination lines are not suitable probes of low density regions. UV resonance lines of high ionization species are the ideal probes of those regions, but unfortunately no UV data are available for this object.

It should be noted that even though the lower limit for the ratio between the equatorial and polar mass loss rates is 20, the ratio of the actual densities is much larger. As shown in Figure~\ref{density} this ratio is about 200 or larger, depending on the adopted value for $A_1$.

The shape of the absorption component of the \ion{H}{1} lines is highly variable, with the blue absorption edge moving from $\approx 400$ to $\approx 800\rm\; km\;s^{-1}$ on very short timescales (see Fig. 3 of Paper III). For this reason, it is difficult to reproduce the detailed shape of the absorption profile. The high variability of the observed line profiles suggests variable mass loss on short timescales. For instance, our models reproduce very well the H$\beta$ absorption profile of 2006, December 27, but not the profile that was observed two weeks earlier. This may indicate that not only the mass loss vary but also the kinematic properties of the CS envelope change as well.

The depth of the absorption component, contrary to its shape, is remarkably constant. It depends on two competing processes: the absorption of resonant radiation by \ion{H}{1} atoms in a given excitation state and  the emission filling-in, i.e., the generation of new photons both by line emission and bound-free and free-free continuous emission that partially fills in the absorption component. The filling-in by continuous emission is particularly important because of the very large densities close to the star (see Figure~\ref{density}). As shown in Fig.~\ref{hghd}, our models also fit reasonably well the depth of the absorption component for other lines of the Balmer series, which is not a trivial result.

Another important parameter is the value of $\beta$ that tells how fast the wind accelerates. We found that the best models for the line profiles are the ones with large values of $\beta$ for all latitudes. 
Low values of $\beta$ (say, $\beta \lesssim 2$) would result in a too broad emission component, due to the large velocities in the inner part of the wind, and also a in sharp absorption edge at high velocities. Both these features are not observed.
Even though in our models it was possible to have different values of $\beta$ for different latitudes, our best fit value was for $\beta=4$ and $A_3=0$, i.e., a constant value of $\beta$.
Again, this is a somewhat uncertain result. For instance, a model with $A_3 = 1/3$, for which $\beta(0)=3$, is also a good model and cannot be ruled out from the present analysis.

We end this section discussing the temperature and ionization structure of the CS gas. Fig.~\ref{temperature} shows the midplane temperature and the fractional number of H atoms in the ground level as a function of the distance from the star. 
Clearly, the gas in highly non-isothermal, a result already obtained by \citet{dre85} and, more recently, by of \citet{car06a} and \citet{zsa08}. 
The temperature reaches a minimum of about $5\,000\;\rm K$ at a distance of around $5\;R$ and then steadily rises to an equilibrium temperature of about  $12\,000\;\rm K$ for $r \gtrsim 100\;R$. This rise in the temperature is probably a result of our unrealistic chemical composition. The inclusion of the cooling terms by other elements will result in a different temperature profile \citep[see, for instance,][]{dre85}.

The gas is essentially ionized everywhere in the envelope. The maximum value for the relative fraction of \ion{H}{1} and \ion{H}{2} is never larger than 10\%. This result agrees with the recent work by \citet{zsa08}, who studied the envelopes of sgB[e], and found that H is generally ionized, although  this may depend on both the effective temperature of the central star and on the density scale of the wind.

\subsection{Circumstellar Dust \label{CSdust}}

\begin{figure}
\centerline{\includegraphics[width=8cm]{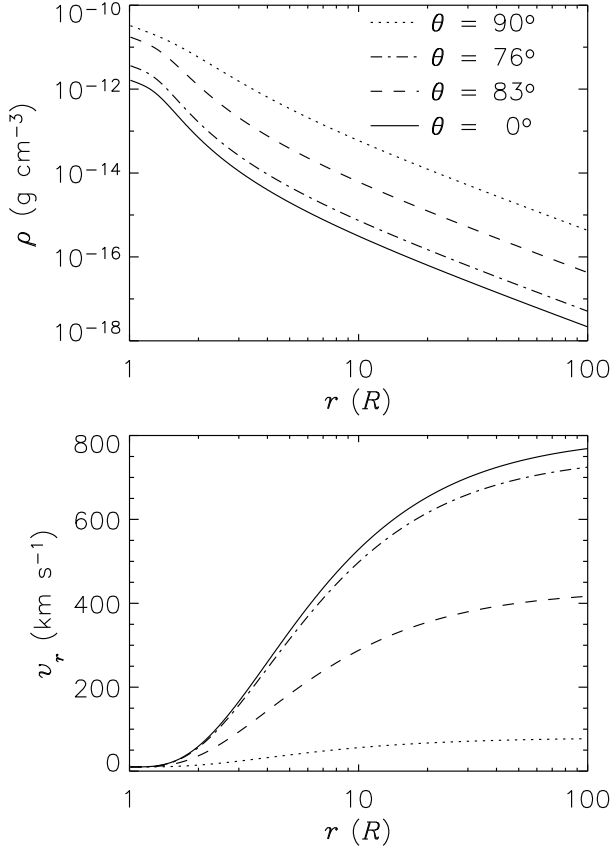}}
\caption{
Top: gas density vs. distance to the star, for several colatitudes, as indicated. $\theta=90\degr$ corresponds to the equator.
Bottom: radial velocity vs. distance.
\label{density}}
\end{figure}

\begin{figure}
\centerline{\includegraphics[width=8cm]{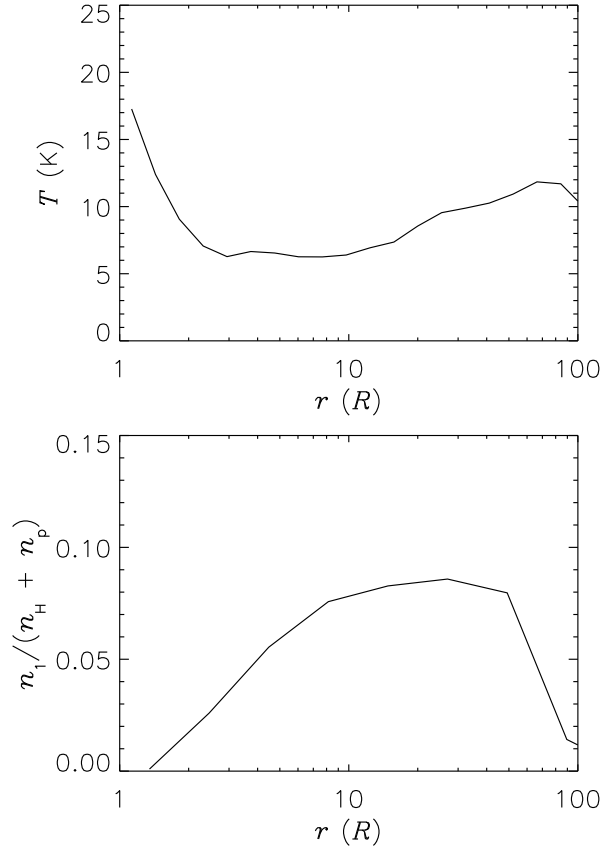}}
\caption{
Top: electron temperature along the midplane vs. distance to the star.
Bottom: ratio between the number density of H atoms in the ground level, $n_1$, to the total number of H atoms ($n_{\rm H}$) and protons ($n_{\rm p}$).
Shown are the results for $d\dot{M}(90\degr)/d\Omega=6\times10^{-9}\;M_\sun \rm\;yr^{-1}\;sr^{-1}$.
\label{temperature}}
\end{figure}

\begin{figure}
\includegraphics[height=6.5cm]{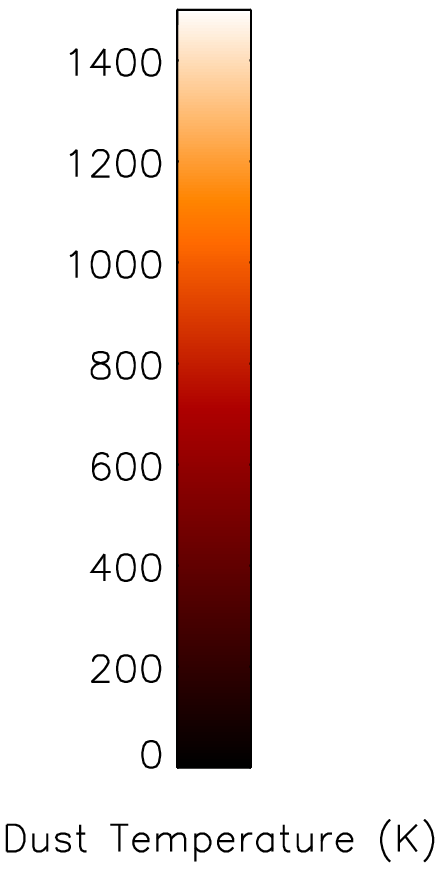}
\includegraphics[height=6.5cm]{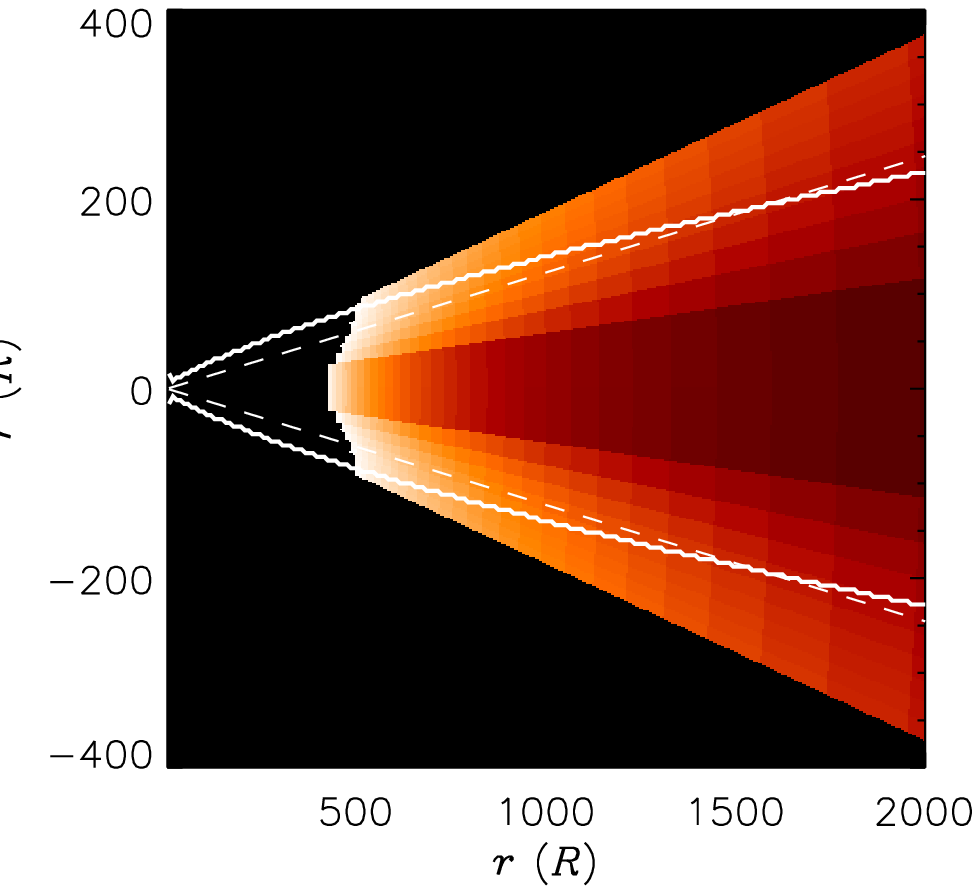}
\caption{
Dust grain temperatures for the smallest grains of dust model 1 ($a=0.05\;\mu\rm m$).
The dashed lines indicate the disk opening angle.
The solid lines enclosure the region of the disk for which eq.~(\ref{rho_crit2}) is valid. A value of $\eta=0.2$ was used.
Note that the figure is not in scale, the $y$-axis was enlarged to bring to evidence the details of the geometrically thin dust destruction surface.
\label{dust2d}}
\end{figure}

\begin{figure}
\centerline{\includegraphics[width=8cm]{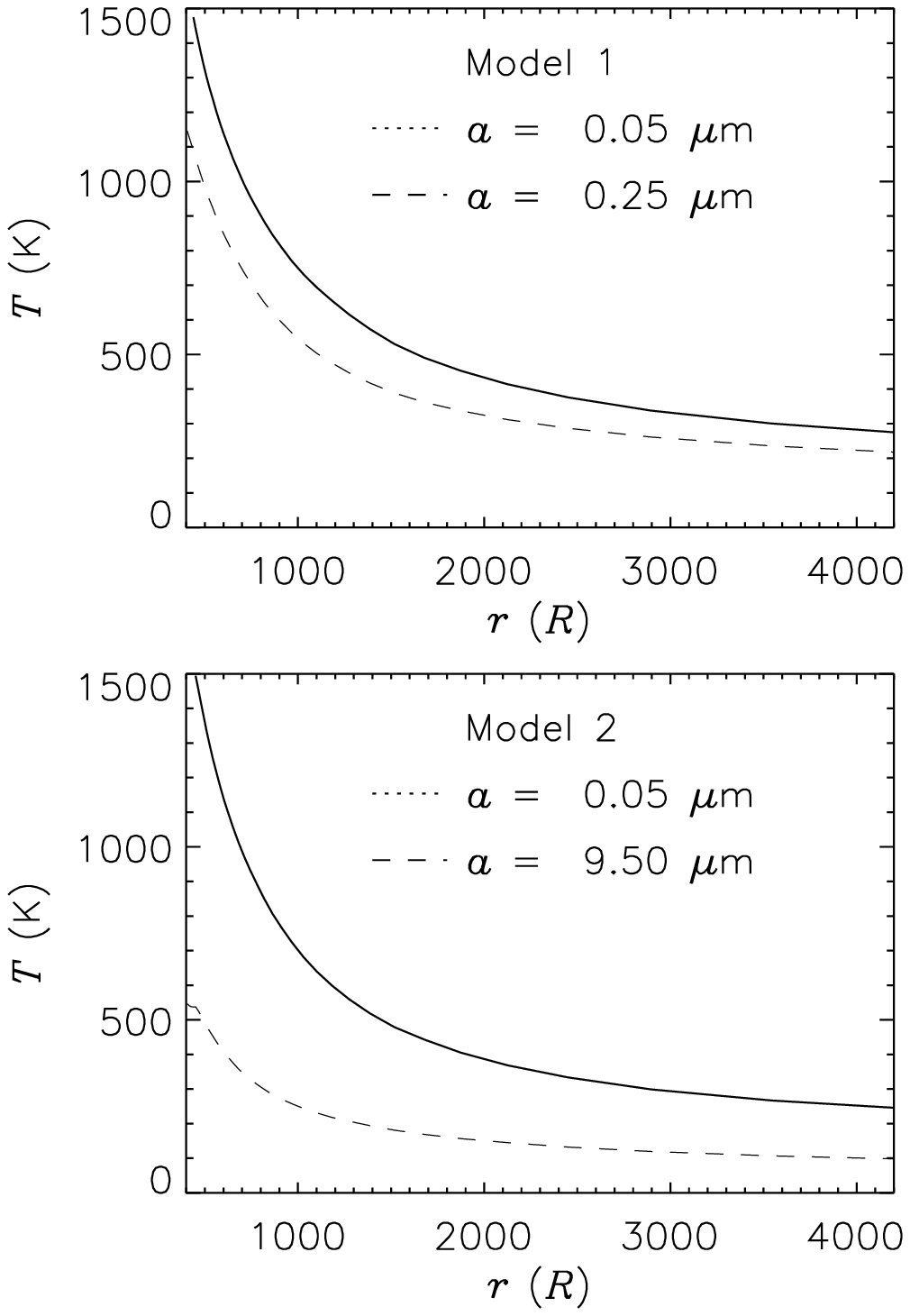}}
\caption{
Dust equilibrium temperature as a function of distance to the star.
Top: dust model 1. Bottom: dust model 2. In each panel we show the temperature for the smallest (solid line) and largest (dashed line) grains of the distribution.
\label{temp_vs_a}}
\end{figure}

Many processes are involved in defining the shape of the SED, and both gas and dust can play an important role depending on the wavelength range considered.
In the visible region, the dominant processes are continuum absorption by both gas and dust, and emission by continuum processes in the gas (bound-free and free-free emission). 
The SED is controlled by both gas and dust emission at wavelengths of $\lesssim 2\;\mu\rm m$ in the near-IR, but the dust becomes the major player at longer wavelengths

For this initial study of the dusty content of IRAS\,00470+6429 three different models for the CS dust were investigated. 
For all models the grain size distribution has the form of eq.~(\ref{MRN}) with $n=-3.5$, and each model explores different values for the minimum and maximum sizes of the distribution.
The first model studied corresponds to the standard MRN model for interstellar dust grains \citep{mat77}, for which $a_{\rm min}=0.05$ and $a_{\rm max}=0.25\;\mu\rm m$. The second model employs a much larger upper limit for the distribution ($a_{\rm max}=10\;\mu\rm m$), whereas the third model has very large grains only ($a_{\rm min}=1$ and $a_{\rm max}=50\;\mu\rm m$).
The three dust models are listed in Table~\ref{t6}.

At this point, it is useful to consider what controls the shape and level of the IR excess arising from dust thermal emission. The {shape} of the emitted spectrum is controlled mainly by the dust absorption coefficient in the IR domain, which is a function of chemical composition and size of the grain. Other factors controlling the shape are the temperature and density structure.
The {level} of the IR excess depends on the fraction of the radiation coming from the central regions that is reprocessed by the dust grains. 
Since in our model the dust is confined to the dense equatorial outflow, the amount of reprocessing will be a function of the disk opening angle, $\Delta\theta_{\rm op}$, and the dust optical depth, which is set by the gas-do-dust ratio and $\rho_{\rm dust}$ (the bulk density of the grain material).

The role of $\eta$ in defining the dust condensation site deserves a more careful examination.
Recall that the dust condensation site is defined by two conditions: the density of the growing material (Si) must satisfy eq.~(\ref{rho_crit2}), in which $\eta$ is a scaling factor, and the dust equilibrium temperature must be smaller than a given value that corresponds to the temperature above which the dust grains are destroyed (Sect.~\ref{dustrecipe}). In Fig.~\ref{dust2d}, we plot the equilibrium temperature of the smallest grains considered by us ($a=0.05\;\mu\rm m$). As we see below, the smallest grains are always hotter than the larger grains, and, therefore, are used as the tracers of the dust formation site.
The plot represents a cut through the envelope such that the direction of the symmetry axis is vertical and the direction of the midplane is horizontal.
The latitude-dependent inner dust radius represents our solution for the dust destruction surface. 
Near the midplane, where the densities are larger and the material is more shielded from the central radiation, the dust destruction surface is closer to the star, and this surface moves outward for higher latitudes. 

The solid white lines in Fig.~\ref{dust2d} mark the points of the wind for which the right side of eq.~(\ref{rho_crit2}) is equal to the left side, and it encloses the region around the equator which have densities larger than the critical density for grain formation (to draw this line we used $\eta=0.02$, see below). According to our two assumptions, the site for dust formation is the region of the wind both beyond the destruction surface and inside the region enclosed by the solid white lines.
The dust condensation site has an angular size of $2\Delta\theta_{\rm dust}$, the value of which is controlled by the parameter $\eta$: the larger $\eta$ the smaller $\Delta\theta_{\rm dust}$.

It is important to note that the site for dust formation does not necessarily coincides with the site where dust actually is, because the dust formed in the innermost region of the condensation site will be carried outwards along the wind streamlines. 
Therefore, dust that was formed at high latitudes will be dragged to regions where the density is much too low to form dust.
This mechanism allows dust to exist at high latitudes in the outer envelope, where the density is lower than the critical value.
Therefore, in our models dust is confined within a wedge with a (half) opening angle of $\Delta\theta_{\rm dust}$. $\Delta\theta_{\rm dust}$ is a function of both $\Delta\theta_{\rm op}$ and $\eta$.

Figure~\ref{sed1} compares the best-fit SED obtained for each of the three dust models tested.
We did not attempt to fit the IRAS fluxes, but focus instead on the BASS and MSX data, since the large aperture of IRAS has probably lead to some degree of contamination by the interstellar cirrus. Also, we recall, from the data description in Paper III, that the NIRIS data have large errors of up to 20\% in the photometric calibration.

The MRN model (model 1, top panel of Fig.~\ref{sed1}) does not fit well the data. 
The integrated IR flux is in rough agreement with the observations, which means that the model does reprocess the correct fraction of the bolometric stellar luminosity. It reproduces the SED reasonably well in the 2 --- $3\;\mu\rm m$ range, but fails to reproduce the slope of the SED for larger wavelengths.
The model with a large upper limit for the grain size  (model 2, middle panel of Fig.~\ref{sed1}) fits both the level and shape of the SED in the 2 --- $7\;\mu\rm m$ range. 
The reason why model 2 fits better the shape of the near-IR SED than model 1 resides in an interesting physical effect:
differently sized grains have different equilibrium temperatures \citep{car04}.
Figure~\ref{temp_vs_a} shows the temperature of the smallest and largest grains of dust models 1 and 2.
For Model 1, the span in grain sizes is not very large and the difference between 
the coolest and hottest grains is, at most, $300\;\rm K$.
For Model 2, however, we find that the very large grains are much cooler then the small grains.
This explains the difference in the slope of the SED: model 2, having on average a lower dust equilibrium temperature, have proportionally larger emission at longer wavelengths.

The $10\;\mu\rm m$ region of the spectrum presents a big challenge for models 1 and 2, since for both the $9.7\;\mu\rm m$ silicate emission feature is quite strong, in sharp contrast to the observations.
This situation is much improved for the model with only large grains (model 3, bottom panel of Fig.~\ref{sed1}). This model reproduces reasonably well the shape of the thermal IR emission in the entire 2 --- $13\;\mu\rm m$ range. 
As discussed above, the shape of the IR thermal continuum is controlled mainly by the dust absorption opacity, which is a function of grain size and composition. Under our assumption that the grains in IRAS\,00470+6429 are oxygen-based, the observed IR excess seems to indicate that large grains ($a \gtrsim 1\;\mu\rm m$) are the dominant constituents of the CS dust.

The large difference between the equilibrium temperatures of small and large grains have another consequence for the models. Recall that one of the conditions for determining the dust condensation site is that the temperature of the hottest grains cannot exceed $T_{\rm destruction}$.
For models 1 and 2 this condition is satisfied only for $r \gtrsim 540\;R$, but for model 3 the dust destruction surface is much closer to the star, $r \gtrsim 79\;R$.

Another important difference between models with small and large grains lies in the bulk density of the grain material ($\rho_{\rm dust}$). At a given location of the envelope, a constant fraction of the gas is converted into dust. Since the mass of the grain increases as $a^3$ and its cross section as $\sim a^2$, we had to decrease the bulk density of the material so that the dust optical depth remains approximately constant. 
Therefore, our modeling for the IR excess of IRAS\,00470+6429 suggests that the dust grains are both large ($a \gtrsim 1\;\mu\rm m$) and fluffy ($\rho_{\rm dust} \sim 0.1\rm\; g \;cm^{-3}$).

Finally, the three models also reproduce well the visible SED (Fig.~\ref{sed1}), but require different amounts of interstellar reddening: for the models with small grains,  $E(B-V)_{\rm IS}=1.1$, and for the model with large grains,  $E(B-V)_{\rm IS}=1.2$. Recall that $E(B-V)_{\rm IS}$ is a free parameter in the modeling and was set by fitting the visible SED. The different values of  $E(B-V)_{\rm IS}$ can be easily understood by considering that the CS extinction of model 3 is essentially gray in the visible and for this reason its intrinsic CS reddening is smaller than the models with small grains.

The extent to which the CS disk modifies the stellar radiation can be assessed by comparing the model SED with the unattenuated stellar radiation (dashed line in Fig.~\ref{sed1}). 
Since the star is viewed close to edge on, this causes the extinction of UV and visible radiation, that is reprocessed into longer wavelengths. Most of the reprocessed radiation escapes in the polar direction, due to the fact that the vertical optical depth is much smaller than the radial optical depth, but part of it escapes radially. 
The $U$-band flux is particularly important for the SED fitting.  
Our models predict that when the object is viewed pole-on, the large bound-free excess produced by the inner envelope  completely fills in the photospheric Balmer jump, thus producing a power-law SED up to the UV region. The fact that the Balmer jump is present is another evidence in support of larger inclination angles (Sect.~\ref{CSgas}).

\section{Discussion}\label{discus}

At this point it is useful to recall the assumptions of the model presented here.
For this exploratory study it was assumed a bimodal wind composed of a slowly outflowing disk and a fast polar wind, and an ad hoc density and velocity law was used to explore different characteristics of this bimodal wind model, such as the ratio between the mass loss rate at the pole and the equator, and the opening angle of the dense equatorial outflow. Here we intentionally limited the number of free parameters and focussed our analysis on the CS region, aiming at studying its gaseous and dusty phases.
The use of the radiative transfer code \hdust allowed us to perform a fully consistent solution of the radiative transfer problem. Since the results are not marred by unknown radiative transfer effects, the comparisons between the models and observations represent genuine tests of the physical assumptions.

In our models the line emission comes mostly from the dense outflowing disk, which has an opening angle of $\Delta\theta_{\rm op} = 7\degr$, whereas  the strong line absorption component originates in the upper layers of the disk.
The model line profiles reproduce the emission component of the  \ion{H}{1} lines, as well as important features of the absorption component, such as its depth, reasonably well. 

The total stellar mass loss can be calculated by integrating eq.~(\ref{mdot}) over all solid angles
\begin{equation}
\dot{M} = 2\pi \int_0^{\pi}\frac{d\dot{M}}{d\Omega}(\theta) \sin(\theta) d\theta\;.
\end{equation}
Using the best-fit values from Table~\ref{t6}, we obtain $\dot{M} = (2.5$ --- $2.9)\times 10^{-7}\;M_\sun\rm\; yr^{-1}$. The lower limit corresponds to the fit for the 2006, December 13 epoch, whereas the upper limit 
describes the 2006, December 27 epoch. 
The total mass loss rates derived here are one to two orders of magnitude lower than what is estimated for sgB[e] \citep{zic06}. On the other hand, it is at least two orders of magnitude larger than the mass-loss rate of main-sequence B stars, which is of the order of  $10^{-10}$ --- $10^{-9}\;M_\sun\rm\; yr^{-1}$ \citep{gat81}.
This result, if corroborated by future studies, leads to the question of how mass loss is driven in FS CMa objects. 

An inclination angle $i\approx84\degr$ was obtained from the modeling. Since the system is viewed nearly edge-on and the emission component of the line profile is narrow, the presence of a large rotational component in the velocity field can be excluded, thereby justifying the pure radial expansion assumed here.
It follows that a rotationally supported Keplerian disk, which is one of the scenarios proposed for the disks around sgB[e] \citep{por03}, can be excluded for IRAS\,00470+6429.

According to our assumptions for grain formation, the dust around IRAS\,00470+6429 is contained within a wedge of opening angle $\Delta\theta_\mathrm{dust} \approx 10\degr$. 
This opening angle was found by fitting the IR excess, which is a direct measure of the amount of radiation that is reprocessed by the CS dust. 
This value of $\Delta\theta_\mathrm{dust}$ implies that the scaling factor $\eta$ in eq.~(\ref{rho_crit2}) is about 0.02, therefore much less than unity. The physical implication of this value of $\eta$ is that the critical density for grain formation derived by \citet{gai88} might be overestimated, i.e., grains might be able to form on shorter timescales.


The models clearly indicate a prevalence of large grains in the dust around IRAS\,00470+6429, for the reason that model 3 was the only model to satisfactorily fit the IR excess and to explain the absence of a conspicuous $9.7\;\mu\rm m$ silicate emission feature, which is to be expected for oxygen-based dust grains.
This result seems to be in contradiction with the usual view on grain formation, according to which large grains form from the coalescence of smaller grains. 
This apparent contradiction can be explained by assuming that the medium around IRAS\,00470+6429 is clumpy.  In such a medium, dense clumps are the ideal sites for the grain formation and growth, because they provide the necessary shielding from the surrounding radiation.
As the clumps expand, following the wind streamlines, they become optically thiner, exposing the grains to more radiation. Because small grains are hotter than large grains, they will have a larger probability to 
be destroyed.
Clumps are thought to exist in the wind of hot stars \citep[e.g.][]{kud00}, and this simple mechanism of selective grain sublimation may be able to explain the observed trend of a prevalence of large grains in IRAS\,00470+6429. A careful investigation of this hypothesis will be left for a future study.



Another implication from our modeling is that there is a significant CS extinction in the visible and UV regions. This follows from the almost edge-on orientation of the equatorial plane of the CS envelope. As a result, the object looks fainter than without the envelope, which reduces its estimated distance. The distance that follows from our modeling is $d \sim 1.1$ kpc, depending on the dust model. However, the large $E(B-V)_{\rm IS}$ that came from the modeling of the SED in the visible range, 1.1 --- 1.2 mag, is more appropriate for a larger distance ($\sim$ 2 kpc). A similar distance estimate was also suggested in Paper III (2.0$\pm$0.3 kpc), but in that work the CS reddening was not accounted for. This apparent discrepancy between distance and $E(B-V)_{\rm IS}$ may be reconciled by increasing the stellar luminosity. As mentioned in Sect.~\ref{stel_par}, the stellar parameters are only loosely constrained by our modeling, and a higher luminosity (and therefore a larger distance to the star) is certainly plausible.

A critical parameter in our modeling is the inclination angle. Better constraints on this parameter will be put after polarization measurements become available. Modeling the polarization along with the SED and spectral line profiles can be done with \hdust and this will allow us to constrain much better the geometrical properties of the CS envelope, including the inclination angle.

\section{Conclusions}\label{concl}

In this paper we adopted an ad hoc model for IRAS\,00470+6429 which is physically motivated by the current knowledge of the CS envelopes of sgB[e].
The model consists of a $6\;R_\sun$ central star with $T_{\rm eff} = 20\,000\;\rm K$ and $L=5\,100\;L_\sun$, surrounded by a bimodal CS envelope composed of  a dense, slowly outflowing disk-like wind and a fast polar wind.
The radiative transfer calculations were performed with the computer code \hdust. 

In the model the underlying physical reason for the bimodal envelope is that the mass loss is, somehow, enhanced around the equator. From the analysis of the observed SED and \ion{H}{1} line profiles, it was found that the equatorial outflow is at least 200 times denser than the fast polar wind and about 10 times slower. In addition, the opening angle of this slowly outflowing disk (defined as the co-latitude for which the mass loss rate drops to half of its value at the equator) is about $7\degr$. Our analysis firmly exclude both very small ($\lesssim5\degr$) and very large  ($\gtrsim15\degr$) opening angles.

We determined that the integrated stellar mass loss rate is $\dot{M} = 2.5$ --- $2.9\times 10^{-7}\;M_\sun\rm\; yr^{-1}$. The two extremes correspond to fits to observations made in December 2006,  but separated by an interval of two weeks. This gives a quantitative measure of how the mass loss varies in short timescales.
The above value of $\dot{M}$, while much smaller than that of sgB[e], which is of the order of $10^{-6}$ --- $10^{-5}\;M_\sun\rm\; yr^{-1}$, is at least 100 times larger than that of main-sequence stars of spectral type B. 

We adopted a prescription for dust formation based on two complementary criteria. In order to form dust at a given point in the CS envelope, the equilibrium temperature of the dust grains must be smaller than a given grain destruction temperature, $T_{\rm destruction}$, and the gas density must be larger than a critical value. 
Several important radiative transfer effects are considered in our calculations, such as the shielding of the dust by the optically thick inner CS material and the fact that differently sized grains have different equilibrium temperatures. 

To investigate the properties of the dusty content of  IRAS\,00470+6429  we studied dust models with three different grain size distributions, one with the standard MRN distribution ($a=0.05$---$0.25\;\mu\rm m$), one with both small and large grains ($a=0.05$---$10\;\mu\rm m$), and one with only large grains ($a=1$---$50\;\mu\rm m$). The dust was assumed to be oxygen-based (silicates), an assumption supported by evolutionary arguments and our Spitzer observations of FS\,CMa stars.

The only model capable of reproducing the observed IR excess in the entire 2 --- $13\;\mu\rm m$ range was the model with only large grains ($a_{\rm min} = 1\; \mu\rm m$), because the presence of small grains always results in a strong $9.7\;\mu\rm m$ silicate emission, which is not observed.
One consequence of the prevalence of large grains around IRAS\,00470+6429 is that the bulk density of the grain material must be very small ($\rho_{\rm dust} \sim 0.1\rm\; g \;cm^{-3}$).
Therefore, the observed shape of the IR excess seems to firmly indicate that the CS dust grains of IRAS\,00470+6429 are both very large and fluffy.  


\acknowledgements{
This work was supported by FAPESP grant 04/07707-3 and CNPq grant 308985/2009-5 to A.C.C. and NSF  grant AST-0307686 to the University of Toledo (J.E.B.).
This research has made use of the SIMBAD database operated
at CDS, Strasbourg, France as well as of data products from the Two
Micron All Sky Survey, which is a joint project of the University of
Massachusetts and the Infrared Processing and Analysis
Center/California Institute of Technology, funded by the National
Aeronautics and Space Administration and the National Science
Foundation. 
Part of the calculations were carried out on the Itautec High Performance Cluster 
of the Astronomy Department of the IAG/USP, whose purchase was made 
possible by the Brazilian agency FAPESP (grant 2004/08851-0).
A.C.C. is grateful for the help of the technical computing staff of the Astronomy Department of the IAG/USP.
The authors are grateful to the anonymous referee for his/her helpful comments and suggestions.
}


\begin{thebibliography}{}
\bibitem[Carciofi et al.(2004)]{car04} Carciofi, A.~C., Bjorkman, J.~E., \& Magalh{\~a}es, A.~M.\ 2004, \apj, 604, 238
\bibitem[Carciofi \& Bjorkman(2006)]{car06a} Carciofi, A.~C., 
\& Bjorkman, J.~E.\ 2006, \apj, 639, 1081 
\bibitem[Carciofi et al.(2006)]{car06b} Carciofi, A.~C., et 
al.\ 2006, \apj, 652, 1617 
\bibitem[Carciofi et al.(2007)]{car07} Carciofi, A.~C., 
Magalh{\~a}es, A.~M., Leister, N.~V., Bjorkman, J.~E., 
\& Levenhagen, R.~S.\ 2007, \apjl, 671, L49 
\bibitem[Carciofi et al.(2008)]{car08a} Carciofi, A.~C., 
Domiciano de Souza, A., Magalh{\~a}es, A.~M., Bjorkman, J.~E., 
\& Vakili, F.\ 2008, \apjl, 676, L41  
\bibitem[Carciofi \& Bjorkman(2008)]{car08b} Carciofi, A.~C., 
\& Bjorkman, J.~E.\ 2008, \apj, 684, 1374
\bibitem[Carciofi et 
al.(2009)]{car09} Carciofi, A.~C., Okazaki, A.~T., Le Bouquin, J.-B., {\v S}tefl, S., Rivinius, T., Baade, D., Bjorkman, J.~E., \& Hummel, C.~A.\ 2009, \aap, 504, 915 
\bibitem[Castor et al.(1975)]{cas75} Cartor, J., Abbott, D. C., \& Klein, R. I. 1975, \apj, 195, 157Tjer
\bibitem[Claret(2000)]{cla00} Claret, A.\ 2000, \aap, 363, 1081 
\bibitem[Clark et al.(2000)]{clark00} Clark, J.S., et al. 2000, \aap, 356, 50
\bibitem[Drew(1985)]{dre85} Drew, J. E. 1985, \mnras, 217, 867
\bibitem[Gail \& Sedlmayr(1988)]{gai88} Gail, H.-P., \& Sedlmayr, E.\ 1988, \aap, 206, 153 
\bibitem[Gathier et al.(1981)]{gat81} Gathier, R., Lamers, 
H.~J.~G.~L.~M., \& Snow, T.~P.\ 1981, \apj, 247, 173 
\bibitem[Grady et al.(1987)]{gra87} Grady, C.~A., Bjorkman, 
K.~S., \& Snow, T.~P.\ 1987, \apj, 320, 376 
\bibitem[Harmanec(1988)]{har88} Harmanec, P.\ 1988, Bulletin 
of the Astronomical Institutes of Czechoslovakia, 39, 329 
\bibitem[Kudritzki 
\& Puls(2000)]{kud00} Kudritzki, R.-P., \& Puls, J.\ 2000, \araa, 38, 613
\bibitem[Kurucz(1994)]{kur94} Kurucz, R.~L.\ 1994, Kurucz CD ROM 19, Solar Model Abundance Model Atmospheres,
 (Cambridge: Smithsonian Astrophysical Observatory)
\bibitem[Lamers 
\& Pauldrach(1991)]{lam91} Lamers, H.~J.~G., \& Pauldrach, A.~W.~A.\ 1991, \aap, 244, L5 
\bibitem[Lamers et 
al.(1998)]{lam98} Lamers, H.~J.~G.~L.~M., Zickgraf, F.-J., de Winter, D., Houziaux, L., \& Zorec, J.\ 1998, \aap, 340, 117 
\bibitem[Lamers 
\& Cassinelli(1999)]{lam99} Lamers, H.~J.~G.~L.~M., \& Cassinelli, J.~P.\ 1999, Introduction to Stellar Winds, Cambridge University Press, Cambridge, UK
\bibitem[Mathis et al.(1977)]{mat77} Mathis, J.~S., Rumpl, 
W., \& Nordsieck, K.~H.\ 1977, \apj, 217, 425 
\bibitem[Miroshnichenko(2007)]{m07} Miroshnichenko, A.S. 2007, \apj, 667, 497
\bibitem[Miroshnichenko et al.(2007)]{m07a} Miroshnichenko, A.S., et al. 2007, \apj, 671, 828
\bibitem[Miroshnichenko et al.(2008)]{m08} Miroshnichenko, A.S., Gray, R.O., Bjorkman, K.S., Rudy, R.J., Lynch, D.K., Carciofi, A.C., \& Men'shchikov, A.B. 2008. BAAS, 40, 2008
\bibitem[Miroshnichenko et al.(2009)]{m09} Miroshnichenko, A.S., et al. 2009, \apj, 700, 209 (Paper III)
\bibitem[Ossenkopt, Henning \& Mathis(1992)]{oss92} Ossenkopt, V., Henning, Th. \& Mathis, J.S. 1992, \aap, 267, 567
\bibitem[Porter(2003)]{por03} Porter, J.~M.\ 2003, \aap, 398, 631 
\bibitem[Tull et al.(1995)]{tull95} Tull, R.G., MacQueen, P.J., Sneden, C., \& Lambert, D.L. 1995, \pasp, 107, 251
\bibitem[Zickgraf et al.(1985)]{zic85} Zickgraf, F.-J., Wolf, B., Stahl, O., Leitherer, C., \& Klare, G.\ 1985, \aap, 143, 421 
\bibitem[Zickgraf(2006)]{zic06} Zickgraf, F.-J.\ 2006, in ASP Conf. Ser. 355, Stars with the B[e] Phenomenon, ed. M. Kraus, \& A.S. Miroshnichenko, (San Francisco: ASP), 135 
\bibitem[Zsarg\'o et al.(2008)]{zsa08} Zsarg\'o, J., Hillier, D.J., \& Georgiev, L. N. 2008, \aap, 478, 543
\end{thebibliography}
\end{document}